\documentclass[%
 aip,
 amsmath,amssymb,
 reprint,%
]{revtex4-1}
\usepackage{soul}
\usepackage{xcolor}

\usepackage{graphicx}
\usepackage{dcolumn}
\usepackage{bm}

\usepackage[utf8]{inputenc}
\usepackage[T1]{fontenc}
\usepackage{mathptmx}
\usepackage{etoolbox}

\include{pythonlisting}
\usepackage{algorithm,algpseudocode,float}
\usepackage{longtable}
\usepackage{booktabs}

\newcommand{\eqn}[1]{\begin{align}#1\end{align}}
\newcommand{\bs}[1]{\boldsymbol{#1}}

\newcommand{\modi}[1]{{\color{black} #1}}
\newcommand{\com}[1]{ }
\newcommand{\bor}[1]{ }

\newcommand{\g}{\textbf{G}}
\newcommand{\gs}{\textbf{G}s}
\newcommand{\np}{\textbf{NP}}
\newcommand{\nps}{\textbf{NP}s}

\newcommand{\sca}{\nu}
\newcommand{\pf}{c}

\def\bnabla{\bs{\nabla}}
\def\bomega{\bs{\omega}}
\def\bbf{\bs{f}}

\def\bG{\bs{G}}
\def\bI{\bs{I}}
\def\blambda{\bs{\lambda}}
\def\bM{\bs{M}}
\def\bq{\bs{q}}
\def\br{\bs{r}}
\def\btau{\bs{\tau}}
\def\bu{\bs{u}}

\def\bv{\bs{v}}

\makeatletter
\def\@email#1#2{%
 \endgroup
 \patchcmd{\titleblock@produce}
  {\frontmatter@RRAPformat}
  {\frontmatter@RRAPformat{\produce@RRAP{#1\href{mailto:#2}{#2}}}\frontmatter@RRAPformat}
  {}{}
}%
\makeatother
\begin{document}

\preprint{AIP/123-QED}

\title[Computational modelling of passive transport of functionalized nanoparticles]{Computational modelling of passive transport of functionalized nanoparticles}
\author{Daniela Moreno-Chaparro*}
\email{$^{\dagger}$nmoreno@bcamath.org }

\affiliation{Basque Center for Applied Mathematics, BCAM. Alameda de Mazarredo 14 , Bilbao 48400, Spain}
\affiliation{University of the Basque Country/Euskal Herriko Unibertsitatea. Barrio Sarriena Leioa, 48940, Spain}
\author{Nicolas Moreno $^{\dagger}$}%
 \email{*dmoreno@bcamath.org} 
\affiliation{Basque Center for Applied Mathematics, BCAM. Alameda de Mazarredo 14 , Bilbao 48400, Spain}%

\author{Florencio Balboa Usabiaga}%
\affiliation{Basque Center for Applied Mathematics, BCAM. Alameda de Mazarredo 14 , Bilbao 48400, Spain}%

\author{Marco Ellero}
\affiliation{Basque Center for Applied Mathematics, BCAM. Alameda de Mazarredo 14 , Bilbao 48400, Spain}
\affiliation{IKERBASQUE, Basque Foundation for Science, Calle de Maria Dias de Haro 3, 48013, Bilbao,Spain}%
\affiliation{Zienkiewicz Center for Computational Engineering (ZCCE), Swansea University, Bay Campus, Swansea SA1 8EN, United Kingdom}%


\date{\today}

\begin{abstract}
Functionalized nanoparticles (NPs) are complex objects present in a variety of systems ranging from synthetic grafted nanoparticles to viruses. The morphology and number of the decorating groups can vary widely between systems. Thus, the modelling of functionalized NPs typically considers simplified spherical objects as a first-order approximation. At the nanoscale label, complex hydrodynamic interactions are expected to emerge as the morphological features of the particles change, and they can be further amplified when the NPs are confined or near walls. Direct estimation of these variations can be inferred via diffusion coefficients of the NPs. However, the evaluation of the coefficients requires an improved representations of the NPs morphology to reproduce important features hidden by simplified spherical models. Here, we characterize the passive transport of free and confined functionalized nanoparticles using the Rigid Multi-Blob (RMB) method. The main advantage of RMB is its versatility to approximate the mobility of complex structures at the nanoscale with significant accuracy and reduced computational cost. In particular, we investigate the effect of functional groups distribution, size and morphology over nanoparticle translational and rotational diffusion. We identify that the presence of functional groups significantly affects the rotational diffusion of the nanoparticles, moreover, the morphology of the groups and number induce characteristic mobility reduction compared to non-functionalized nanoparticles. Confined NPs also evidenced important alterations in their diffusivity, with distinctive signatures in the off-diagonal contributions of the rotational diffusion. These results can be exploited in various applications, including biomedical, polymer nanocomposite fabrication, drug delivery, and imaging.
\end{abstract}

\maketitle

\section{\label{sec:level1}functionalized nanoparticles:\protect\\ Introduction}
Nanoparticles (\nps) are \modi{complex structures ubiquitous on many synthetic} and biological system. \nps\ \modi{can be found in a variety of morphologies, ranging from simple shapes (i.e spheres, cubes, ellipsoids) to more complex structures. Moreover, they can exhibits} a disparate number of functional decorations with characteristic morphologies \modi{that determine their functionality}. Examples of these are organelles, viruses, and grafted nanoparticles, \modi{all of them with a size} in the order of 10 to 200nm.\cite{Choueiri2016,Bao2021} Depending on the field, such decorations are typically referred to as functional groups, spikes,\cite{Moreno2022} \modi{grafts,} patches,\cite{Chen2022} or hairs.\cite{Tang2022} For simplicity, here, we use the functional groups (\g) \modi{notation} to \modi{denote any type of decoration on the surface of the \nps\ core.} \modi{In general,} \modi{\nps\ have unique transport features related to the groups morphology and number ($N_G$)}, as illustrates in FIG 1.A. \modi{In this context,} microrheological \modi{techniques} \modi{emerge as powerful tool for \nps\ study and characterization for various applications, including}  biomedical, polymer nanocomposite fabrication, drug delivery, and imaging,\cite{Chancellor2019a} to name a few. \modi{For example, in the field cancer tumours treatment, investigations on the diffusion of different gold \nps\ and liposomes\cite{Shi2019a} through the mucus and cells have been addressed to improve drug delivery}. 

\begin{figure*}[thpb]
\centering
\includegraphics[width=\textwidth]{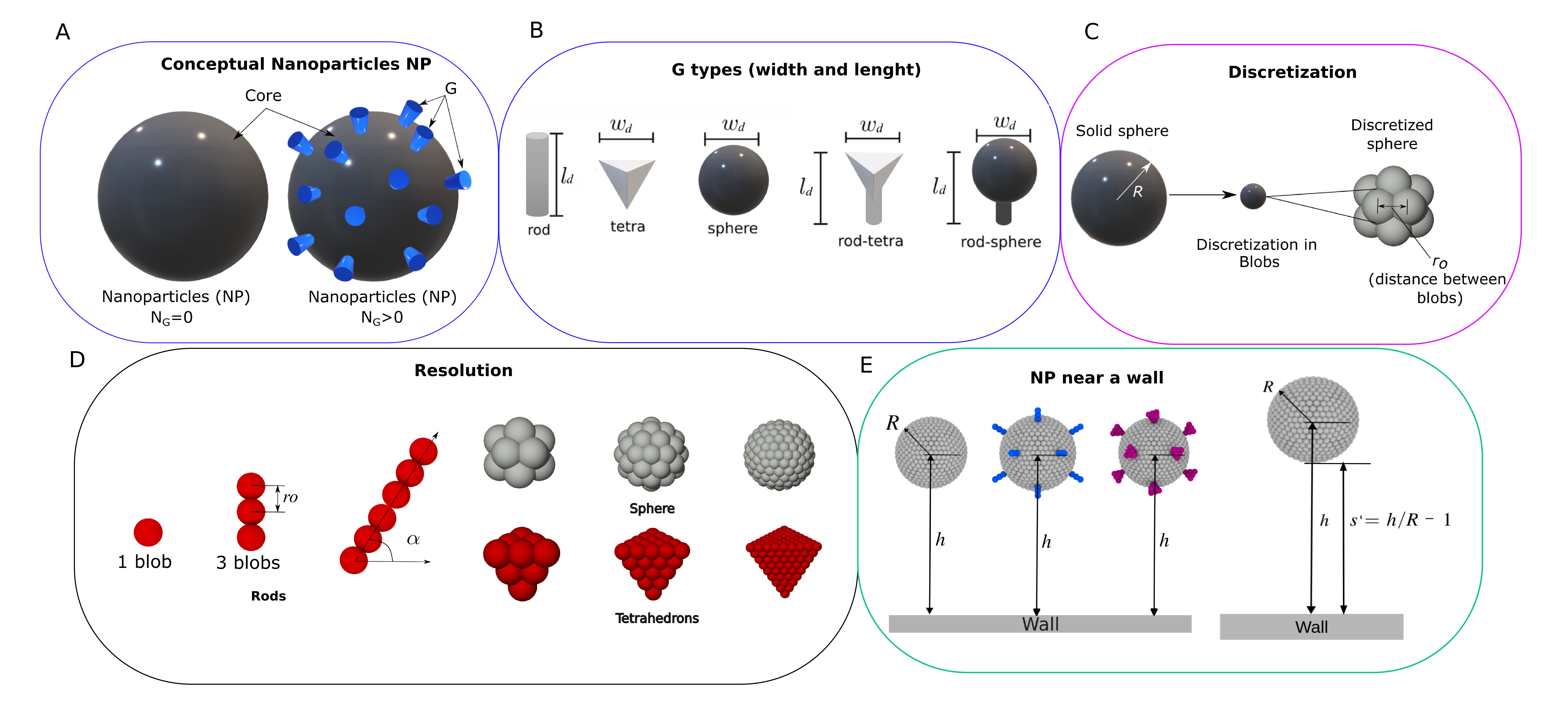}
\caption{\modi{Sketch of a \np\ } and the  discretization adopted. A. \np\ representation with core and functional groups\modi{, \g\,} as blue cylinders. If the number of groups ($N_G$) is $N_G = 0$ the \np\ is only the core. B\modi{.} Functionalized shapes or\g\ types representation, the measurement of length $l_G$ and width $w_G$ are reduced by the radius of the core. C\modi{.} Discretization of a sphere in blobs. D. Resolution and refinement following the\g\ types rod,  tetrahedron, and the core sphere. E\modi{.} \np\ near a wall, the distance h is given by the wall to the center of the \np\ }
\label{fig:tetra}
\end{figure*}

\modi{Computational studies on decorated particles at microscales, have already revealed the effects of \g\ morphology on the formation of complex structures.\cite{Li2014,Karatrantos2017}}\modi{At nanoscales, as thermal fluctuations becomes relevant, investigations has been mostly focused on core-only \nps\ (spheres), whereas the simulation of} functionalized ones remains a challenging task. \modi{The morphological features along with specific binding and affinity interactions between the groups dramatically increase the complexity.} \modi{Different} computational methods have been used to characterize the transport, binding and interactions for \modi{non-functionalized} \nps, \modi{including} Brownian dynamics for free\cite{Li2014,Islam2017a} and near walls \nps,\cite{Lisicki2014} Monte Carlo,\cite{Li2019} coarse\modi{-graining},\cite{Ilnytskyi2020, Debets2020} dissipative particle dynamics,\cite{Liu2012a,Chen2022} and smooth dissipative particle dynamics.\cite{Bian2012,Vazquez2012,Vazquez2022} The \modi{transport} modelling of \modi{\nps\ usually deals with} spherical shapes, \modi{adopting the} Stokes-Einstein equations.\cite{Murray1992} In general, the effect of the groups number, distribution around the core, and morphology, on the \nps\ transport lack of detailed investigations. \modi{Moreover, other relevant aspects} such as their \modi{passive} transport \modi{under} confinement are still missing. \modi{Confined transport may play a critical role in the design} of microfluidic devices\cite{Tang2022} and sensors.\cite{Willner2018}

\modi{Herein}, from a microrheology standpoint, we \modi{investigate} how the mobility of complex functionalized \nps\ is affected by the \g\ morphology, number, and distribution. In particular, we characterize their  translational ($D_t$) and rotational ($D_r$) diffusivity. \np's translational and rotational diffusivity can be  derived using the Stokes-Einstein theory. \modi{\nps\ diffusivity arises from the balance between thermal fluctuations and hydrodynamic interactions with the fluid, that depend on \np\ morphology and fluid viscosity.\cite{Murray1992} Here, we use the Rigid Multi-blob method \cite{BalboaUsabiaga2016} (RMB) to model \modi{complex functionalized \nps\ by discretizing them as a set of rigidly connected spherical blobs}. RMB can be applied to physical and biological systems\cite{Sprinkle2017, Moreno2022, BalboaUsabiaga2022} at the nanoscale, where the thermal fluctuations and hydrodynamic interactions are relevant.} This method, is suitable to model arbitrary shape objects (both free and confined), and its principal advantage  is the low computational cost of solving a mobility problem. \\\\
We consider that the \nps\ are constituted by a spherical core decorated with $N_G$ functional groups. We represent the \np\ core as a spherical shell of multiple blobs, whereas \gs\ are modelled \modi{using various} shapes such as rods, spheres, and tetrahedrons.\modi{These basic set of shapes are inspired from different real morphologies reported for synthetic and biological nanoparticles\cite{Tang2022,Chen2022,Moreno2022,Moreno2020,Choueiri2016}, and allows us to explore in a systematic fashion the effect of \g\ morphology on the nanoparticles transport}. \modi{In general, the number and relative size of the \gs\ can vary widely among physical systems. Thus, we also explore the effects of number density and size (length and width) of the groups.} Additionally, we study the effects of groups distribution around the core. Since, the location of \gs\ may affect the overall mobility of the nanoparticles, understanding the effect of the groups distribution provides relevant information for nanoparticle design and optimization, either to enhance or reduce the diffusion. \modi{As a final compelling aspect on \nps\ transport,} we  \modi{investigate confinement effects, by computing the effective parallel and perpendicular diffusion near walls.}  
\vspace{-0.2 cm}
\subsection{\modi{Functionalized nanoparticles} mobility}

The translational diffusion coefficient of a \np\ is related to its translational mobility,\cite{Einstein1905} and similar arguments can be easily extended to the rotational diffusion.\cite{Kim1991} Thus, the diffusion coefficients, $D_t$ and $D_r$, can be expressed proportional to the mobilities as
\eqn{
  D_t = \frac{k_BT}{3} \mathrm{Tr}\left(\bM_t\right),\;\;\; D_r = \frac{k_BT}{3} \mathrm{Tr} \left(\bM_r\right),
}
 where $k_B T$ is the thermal energy and $\mathrm{Tr}$ denotes the trace operator. The mobility components yield the linear and angular velocities of a \np\ ($\bu$ and $\bomega$) in response to applied forces and torques ($\bbf$ and $\btau$),
\eqn{
  \label{eq:mobility_problem}
  \left(\begin{array}{c}
    \bu \\
    \bomega
  \end{array}\right) =
  \left(\begin{array}{cc}
    \bM_t & \bM_c \\
    \bM_c^T & \bM_r
  \end{array}\right)
  \left(\begin{array}{c}
    \bbf \\
    \btau
  \end{array}\right).
}
Theoretically, for a spherical \np\ of radius $R$ translational diffusion is given by \eqn{D_t=k_BT/(6\pi\eta R),} and the rotational diffusion by \eqn{D_r=k_BT/(8\pi\eta R^3).} For \nps, the mobility components can be calculated using the Stokes equations to a good approximation.\cite{Schmidt2004, BalboaUsabiaga2013}
In this limit, the fluid velocity and pressure, $\bv$ and $p$, obey the Stokes equations with viscosity $\eta$
\eqn{
  \label{eq:Stokes}
  -\bnabla p + \eta \bnabla^2 \bv &= 0, \\
  \bnabla \cdot \bv &= 0,
}
while for boundary conditions, one can assume that the fluid velocity obeys the no-slip condition at the \nps\ surface and decays to zero at infinity.\\
We have assumed a functionalized \np\ behaves like a rigid body; thus, the no-slip condition for a \np\ located at free point $\bq$ is quite simple,
\eqn{
  \label{eq:no-slip}
  \bv(\br) = \bu + \bomega \times (\br - \bq)\;\;\; \mbox{for all } \br \in \partial \Omega.
}
These partial differential equations are closed by the balance of force and torque. The integral of the fluid traction, $-\blambda$, over the surface of the \np\ balance the external forces and torques applied to the \np\ \cite{Pozrikidis1992}
\eqn{
  \label{eq:balance_f}
  \int_{\partial \Omega} \blambda\, \mathrm{d} S_r &= \bbf, \\
  \label{eq:balance_tau}
  \int_{\partial \Omega} (\br - \bq) \times \blambda\, \mathrm{d} S_r &= \btau.
}
In most applications \nps\ are under different type of confinements. Thus, two paradigmatic cases are \textit{i)} \nps\ immersed in a suspension and \textit{ii)} a single \np\ diffusing near a large flat wall. The first case can be modeled by using a computational domain with periodic boundary conditions. In this case the diffusion coefficient is given by\cite{Bian2012}
\begin{equation}
    D = D_t/\lambda
\end{equation}
where $D_t$ is the theoretical translational diffusion and $\lambda$ is a drag coefficient 
\begin{equation}
\centering
    \lambda = (1-1.7601C^{1/3}+C-1.5593C^2+3.9799C^{3/8}
\end{equation}
$$ -3.073C^{10/3}+C^{11/3})^{-1}, $$ where $C=4/3\pi R^3/L^3$ and $L$ is the size of the domain.

In the second case, a \np\ diffusing near a wall, the symmetry of the system is broken by the presence of the boundary. Therefore, it is necessary to distinguish between the diffusion parallel and perpendicular to the wall. If we denote as h the distance from the centre of the NP to the wall, and s=h-R the distance from the NP's core to the wall, the parallel diffusion is given by $D_\parallel(s)=D/\lambda_\parallel(s)$, and the perpendicular is $D_\perp(s)=D/\lambda_\perp(s)$.\cite{Bian2012} Theoretical values of the drag coefficients $\lambda_\parallel(s)$ \cite{Hasimoto1959,Swan2007,Bian2012} and $\lambda_\perp(s)$\cite{Hasimoto1959,Swan2007,Bian2012} are 
\begin{equation}
    \lambda_\perp(s) = \frac{4}{3}\sinh\alpha \sum^\infty_{n=1}\frac{n(n+1)}{(2n-1)(2n+3)}
\end{equation}
$$\left[\frac{2\sinh(2n+1)\alpha+(2n+1)\sinh2\alpha}{4\sinh^2(n+1/2)\alpha-(2n+1)^2\sinh^2\alpha}-1\right],$$
\begin{equation}
    \lambda_\parallel (s)= \left[1-\frac{9}{16}\beta+\frac{1}{8}\beta^3-\frac{45}{256}\beta^4-\frac{1}{16}\beta^5\right]^{-1},
\end{equation} where $ \alpha = \cosh^{-1}(1-h/R)$ and $\beta=(1-h/R)^{-1}$ respectively.

\subsection{Rigid Multi-Blob method (RMB)}

To compute the mobilities of a NP, we adopt the Rigid Multi-Blob method (RMB).\cite{BalboaUsabiaga2016} We discretize the surface of the \np\ with $N$ markers or \emph{blobs} of radius $a$ with position $\br_i$. The blobs are subject to constraint forces, $\blambda_i$, that ensure the rigid motion of the whole NP. Evaluating the no-slip condition at the blobs, as in collocation methods, leads to a linear system of equations for the unknowns $\bu$, $\bomega$ and $\blambda_i$,
\eqn{
  \label{eq:no-slip_blobs}
  \bv(\br_i) = \sum_{j=1}^N \left(\bM_B \right)_{ij} \blambda_j &= \bu + \bomega \times (\br_i - \bq)\;\;\; \mbox{for } i=1,\ldots,N,\\
  \label{eq:balance_f_blobs}
  \sum_{j=1}^N \blambda_j &= \bbf, \\
  \label{eq:balance_tau_blobs}
  \sum_{j=1}^N (\br_j - \bq) \times \blambda_j &= \btau,
}
In the no-slip equation, Eq. \eqref{eq:no-slip_blobs}, the blob mobility matrix $\left(\bM_B\right)_{ij}$
couples the force acting on the blob $j$ to the flow generated at the blob $i$. We use the regularized Rotne-Prager mobility, that has closed analytical expression\cite{Rotne1969,Wajnryb2013}
\eqn{
  \left(\bM_B\right)_{ij} = \left(\bI + \frac{a^2}{6} \bnabla^2_{\br}\right) \left(\bI + \frac{a^2}{6} \bnabla^2_{\br'}\right)  \bG(\br, \br')
  \vert^{\br=\br_i}_{\br'=\br_j},
}
where $\bG(\br, \br')$ is the Green's function of the Stokes equation (i.e. the Oseen kernel). To model the \np\ near a wall, we use the Rotne-Prager-Blake tensor that accounts for the hydrodynamic interactions with the wall.\cite{Sprinkle2017} For reliable hydrodynamics interactions, we recommend a minimum distance of one blob radius $r_o=a$, such that the blobs do not overlap the wall.
\vspace{-0.5 cm}

\subsection{Morphology of functional groups and core of the nanoparticle }

The morphology of a \np\ is given by its spherical-shape core and the functional groups around it. This functional groups have a characteristic length, $l_G$, and width, $w_G$. To streamline the analysis, we consider the dimensionless size of the groups given by $l_G /R$ and width as $w_G /R$, where $R$ is the radius of the \np's core. For simplicity, we choose three general shapes to construct the functional groups: rods, tetrahedrons, and spheres. Using these general shapes, we create two composite shapes, rod-tetra and rod-sphere, illustrated in Figure 1.B. The position of the groups in the \np\ is modelled using uniform and random distributions. For the uniform case, the \gs\ are distributed at equidistant positions corresponding to the vertex of a icosahedra, whereas for random distribution, any position over the core is allowed.

\vspace{-0.5 cm}
\subsection{Discretization of nanoparticles}

The \nps\ are discretized by blobs rigidly connected. The distance between these blobs is $r_o$ and defines the blob size. In FIG 1.C, we illustrate the discretization of a solid sphere of radius $R$ into 12 connected blobs. The general methodology to discretize the different shapes investigated is as follows: \textit{i)} For rods, the construction consists in equidistantly blobs along an orientation angle $\alpha$ (see FIG 1.D) \textit{ii)} for spheres we start with a coarse 12-blobs model located at a distance $R$ from the center, and arbitrary distance $r$ (see FIG 1.C). Then we conduct a recursive refinement taking the middle point of the segments connecting two adjacent vertexes and projecting those points radially, their new position satisfies $R^2 = x^2+y^2+z^2$. This procedure is repeated until $r\leq r_o$. Spheres and tetrahedrons with different degrees of refinement are illustrated in FIG 1.D. \textit{iii)} For tetrahedron, with follow a similar iterative process, starting with coarse surface with only four vertexes. Then the structure is refined by splitting in half the edges between two vertexes and adding a new point in that position. This addition must be applied to all the edges of the primary surface. This procedure is repeated until the distance between adjacent points is smaller than the target resolution.

\vspace{-0.3 cm}
\subsection{Resolution}
We define the resolution ($\Phi$) of the discretization as the ratio between the radius of \np\ core and the distance between blobs as $\Phi = {R}/{r_o}$. The optimal $\Phi$ is a selected as a trade off between accuracy and computational cost. For simple spheres the accuracy at a given resolution we can estimated comparing the hydrodynamic radius computed numerically with the input radius of the object. For tetrahedron shapes, we can define different characteristic sizes such as the width ($a$), height ($H_{\text{tetra}} = {\sqrt{6}}/{3} a$), and the circumscribing sphere of radius $R_{\text{tetra}} = l_G(3/8)^{1/2}$. However, to streamline the resolution analysis for this non-spherical shapes we use its equivalent radius ($R_{e}$), defined by the radius of a sphere with equivalent volume of the tetrahedron, $V_{tetra} = {a^3}/{6\sqrt{2}}$, leading to $R_{e} = \Big({3}/{4\pi} V_{\text{tetra}}\Big)^{{1}/{3}}.$

\subsection{Reduced translational and rotational diffusivities}
\label{DimensionlessDiff}
For convenience, in the remaining, we discuss our findings in terms of reduced diffusivities. For non-functionalized (only spherical core) nanoparticles, we the theoretical translational and rotational mobilities are given by $\bM_{t}^o={1}/{6\pi \eta R}$ and $\bM_{r}^o={1}/{8\pi \eta R^3}$, respectively. The translational and rotational mobilities computed numerically are  simply referred as $\bM_{t}|_ {\text{sphere}}$ and $\bM_{r}|_ {\text{sphere}}$ according the Eq. 17. Henceforth, we define the reduced diffusivities as the ratio $\bar{D}_i = \bM_{i}|_ {\text{sphere}}/{\bM_{i}^{o}}$ where $i=t,r$. For functionalized \np, since no theoretical values exist, we define the reduced diffusivities in terms of the numerical mobilities of the nanoparticles and the non-functionalized core as $\bar{D}_i = {\bM_{i}}/{\bM_{i}|_ {\text{sphere}}}$. 
To indentify the adequate resolution of tetrahedral shapes used for functional groups, we conduct additional studies of freely diffusing tetrahedrons. For this shape in particular since we do not count with an analytical expresion fo the mobility, we express its reduced diffusivities as $\bar{D}_i^{tetra} = {\bM_{i}|_{\text{tetra}}}/{\bM_{i}^{o}}$, where $i=t,r$ and $\bM_{t}^{o}$ is the mobility of a reference sphere. The diffusivity of confined \np\ is analyzed in terms of the parallel $D\parallel$ and the perpendicular $D\perp$ componets to the confining wall. To investigate the effect of confinement, we place the \nps\ at different distances $s'$ to the wall (see FIG 1.E). We denote the diffusivities near a wall as $D_i|_s$ and the ones of a free \np\ (without a wall) as $D_i^*$, where $i=t,r$. Thus, we define the reduced translational and rotational diffusivities for confined \nps\ as $\overline D_{t}^{wall}={D_{t}|_s}/{D_{t}^*}$ and $\overline D_{r}^{wall}={D_{r}|_s}/{D_{r}^*}$, respectively. In each case, $\overline D_{t}^{wall}$ and $\overline D_{r}^{wall}$ are separated into parallel and perpendicular components as explained in Eqs. 12 and 13.

\section{Discussion and Results}

In this section we present first the resolution tests for general shapes such as spheres (core) and tetrahedrons (functional groups). Here, we introduce the appropriate resolution for each shape's accuracy and computational cost. We also estimate such resolution effects for the whole functionalized \np\ considering the core and functional groups to select the appropriate discretization used in the simulation of \nps. Afterwards, we systematically present the effects of type, size, distribution and number of functional groups. Finally, we discuss the passive transport of the functionalized \np\ near a walls. We compare our results on translation diffusion with existent theoretical models for confined particles, and provide an empirical fitting that describes the variation in rotational diffusion of the \nps. 

\subsection{Resolution}

\subsubsection{Spherical shape resolution}

First, we conduct resolution studies for simple spheres using six different $\Phi$, 0.9 (12 particles), 1.8 (42 particles), 3.6 (162 particles), 7.2 (642 particles), 14.5 (2562 particles) and 29 (10242 particles). From the computed mobilities, we define the error as $\text{Error}_ {D_t}={(\bM_t - \bM_t^o)}/\bM_t^o$ and $\text{Error}_{D_r}=({\bM_r - \bM_r^o})/\bM_r^o$, and in Table 1, we summarize the results. We identify that a resolution of $14.5$ provides a reasonable approximation with errors on the order of $1\%$ in $\bar{D}_t$, and $3\%$ for $\bar{D}_r$, for a spherical core. We must note that aceptable results  with errors of $2\%$ for $\bar{D}_t$, and $6\%$ for $\bar{D}_r$ can be already obtained with lower resolutions ($\Phi=7.2$). Unless otherwise stated, in the remaining we adopt $\Phi=14.5$ to investigate the \nps\ diffusion.

\begin{table}[h]
\caption{Resolution study for a single sphere (non-functionalized core)}
\label{SphereoidRes}
\centering
\begin{tabular}{@{}cccccc@{}}
\toprule
\textbf{Resolution} &
\textbf{\begin{tabular}[c]{@{}c@{}}Number of\\  particles\end{tabular}} &
\textbf{$\bM_t$} &
\textbf{$\bM_r$}  &
\textbf{$\frac{\bM_t - \bM_t^o}{\bM_t^o}$} &
\textbf{$\frac{\bM_r - \bM_r^o}{\bM_r^o}$} \\ \midrule
\multicolumn{6}{c}{} \\
29   & 10242 & 0.0527 & 0.0391 & 0.005 & 0.016 \\
14.5 & 2562  & 0.0524 & 0.0384 & 0.012 & 0.035 \\ 
7.2  & 642   & 0.0518 & 0.0372 & 0.024 & 0.065 \\
3.6  & 162   & 0.0503 & 0.0304 & 0.052 & 0.234 \\
1.8  & 42    & 0.0472 & 0.0297 & 0.110 & 0.254 \\
0.9  & 12    & 0.0420 & 0.0213 & 0.208 & 0.465 \\ \bottomrule
\end{tabular}
\end{table}

\vspace{-0.3 cm}
\subsubsection{Tetrahedral shape resolution} 

The discretization of functional groups can define the overall resolution of the \np\ as they are the smallest morphology to be resolved. However, for typical nanoparticles with  $R>l_G$, the number of blobs to discretize the core will be significantly larger than the minimum required for optimal accuracy. Therefore, we focus on finding the  minimal resolution needed to model these functional groups up to a good approximation. In table 2, we present the computed mobilities for different resolutions $\Phi = R_{\text{tetra}}/r_o$, and the convergence criteria for each case. For these shapes, we use the relative difference of the computed mobilities with the one for the highest resolution simulated ($\Phi=40$). For the resolution of $\Phi=2.5$, we obtain an error of $3\%$ for $M_t$ and $8\%$ for $M_r$, further improvement is achieved with $\Phi=4.9$ with $M_t$ and $M_r$ errors of $1\%$ and $4\%$ respectively. Higher resolutions decrease further the error. However, the associated cost to model such a level of refinement significantly increases. In general, for practical reasons, we identify that values of $\Phi=2.5$ provided a reasonable approximation for tetrahedral shapes.

\vspace{-0.3 cm}
\begin{table}[!ht] 
  \renewcommand\thetable{2}
\centering
\caption{Resolutions of a discretized tetrahedron. The mobility of a sphere of radius $R_{e.t}$ for a solid tetrahedron of edge $a$ reduces the mobilities for the different resolutions. As a convergence estimator, we measure variation in the reduced mobility concerning the largest resolution simulated.}
\label{TetraResolutionTable}
\begin{tabular}{@{}ccccccc@{}}
\toprule
\text{Resolution} &
  \text{$\bM_t/\bM_t^o$} &
  \text{$\bM_r/\bM_r^o$} &
  \text{\begin{tabular}[c]{@{}c@{}}$R_{e.t}$\end{tabular}} &
  $a$ & {$\frac{\bM_{t \text{max}}-\bM_t}{\bM_{t \text{max}}}$} & {$\frac{\bM_{r \text{max}}-\bM_r}{\bM_{r \text{max}}}$} \\ \midrule
\multicolumn{1}{c}{Tetrahedron} &   --     &    --    & 0.49& 1.63 & 0 & 0 \\
\multicolumn{1}{c}{40} & 0.82 & 0.48 & 0.50 & 1.66 & 0 & 0 \\
\multicolumn{1}{c}{19.6}& 0.82 & 0.47 & 0.51 & 1.68 & 0.002 & 0.005\\
\multicolumn{1}{c}{9.8} & 0.82 & 0.46 & 0.53 & 1.73 & 0.007 & 0.019 \\
\multicolumn{1}{c}{4.9}  & 0.81 & 0.45 & 0.55 & 1.84  & 0.017 & 0.044\\
\multicolumn{1}{c}{2.5} & 0.79 & 0.44 & 0.62 & 2.04 & 0.036 & 0.080\\
\multicolumn{1}{c}{1.2}  & 0.76 & 0.42 & 0.74 & 2.45 & 0.069 & 0.107 \\ \bottomrule
\end{tabular}
\end{table}
\vspace{-0.1 cm}

\subsubsection{Functionalized nanoparticles resolution}\label{CovidResol}

Based on the previous results, we now explore four different resolutions ($\Phi = [14.5, 7.2, 3.6, 1.8]$) for the whole functionalized \nps. In FIG 2, we present the variation in the error for two type of nanoparticles, corresponding to the group types with lowest (rod) and highest (tetra) volumetric fraction. In this case, to verify the resolution we evaluate the mobility difference between the functionalized nanoparticle $M_{NP}$ and the single core $M_s$ (at the same resolution,  $M_s-M_{NP}/M_s$. In general, we identify that at a resolution of $\Phi=7.2$ the change in mobility is already captured with a reasonable approximation. For coarser resolution ($\Phi=1.8$), the characteristic shape and aspect ratio of the group is smeared into a single blob, thus the representation of the groups is not properly accounted. 

\vspace{-0.1 cm}
\begin{figure}[!ht]\label{randomgraphic}
\centering
{\includegraphics[scale = 0.18]{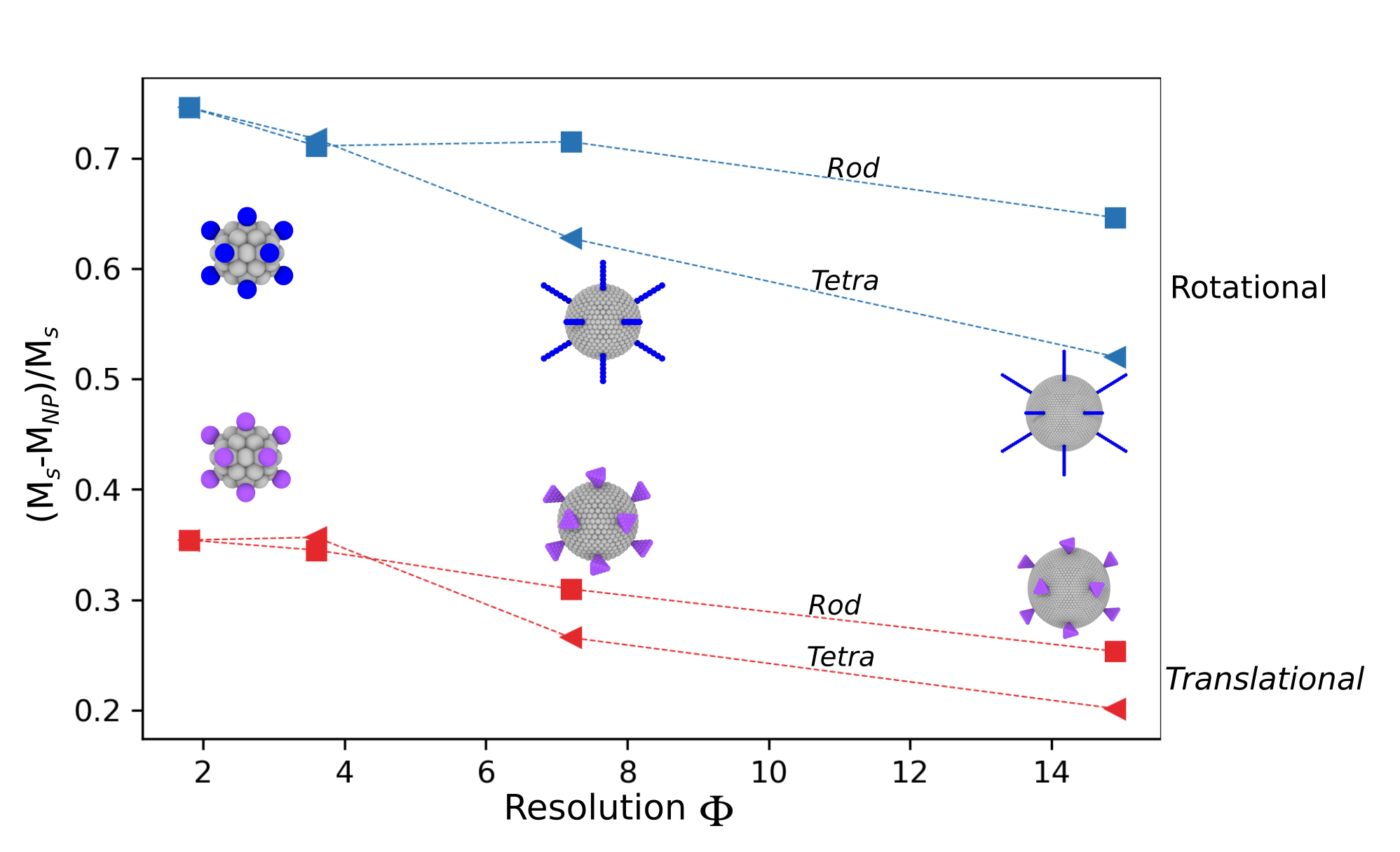}}
\caption{Resolution test for functionalized \nps\ with rods ($l_G/R=1$) and tetrahedron-shaped ($w_G/R=0.2$) groups. The error corresponds to the relative difference in mobilities between the functionalized \np\ and the single spherical core, at the same resolution. The nanoparticles have $N_G=12$ uniformly distributed}
\end{figure}

\subsection{Functional groups morphology and size}

Given the variety functionalized nanoparticules that can be found in the literature, we first investigate how the diffusion of \nps\ with similar number of groups can be affected by the shape and size of the groups. We compute the reduced diffusivities $\overline{D}_t $ and $\overline{D}_r$ of functionalized \nps\ with five group types uniformly placed around on the nanoparticle's surface, as depicted in FIG 3. For consistency, we compare \nps\ with groups of equivalent size $l_G/R=0.5$, $N_G=12$ and uniformly distributed. In general, the presence of the groups induced a reduction in the \nps\ transport due to the effective larger volume. However, this reduction does not occur in a trivial fashion based only on the aspect ratio of the groups. Indeed, the groups shape affects the diffusion due to the added morphological complexity. Overall, the effect of groups type on the translation and rotation differs due to the scaling of $\overline{D}_t \propto R$ and $\overline{D}_r \propto R^3$ with the radius. For example, the reduction in $\overline{D}_t$ for rod and spherical groups is nearly similar, despite their despair effective volume. In contrast, $\overline{D}_r$ appears as a more distinctive parameter to discern \nps\ morphology.

\begin{figure}[!ht]
\centering
\includegraphics[scale = 0.2]{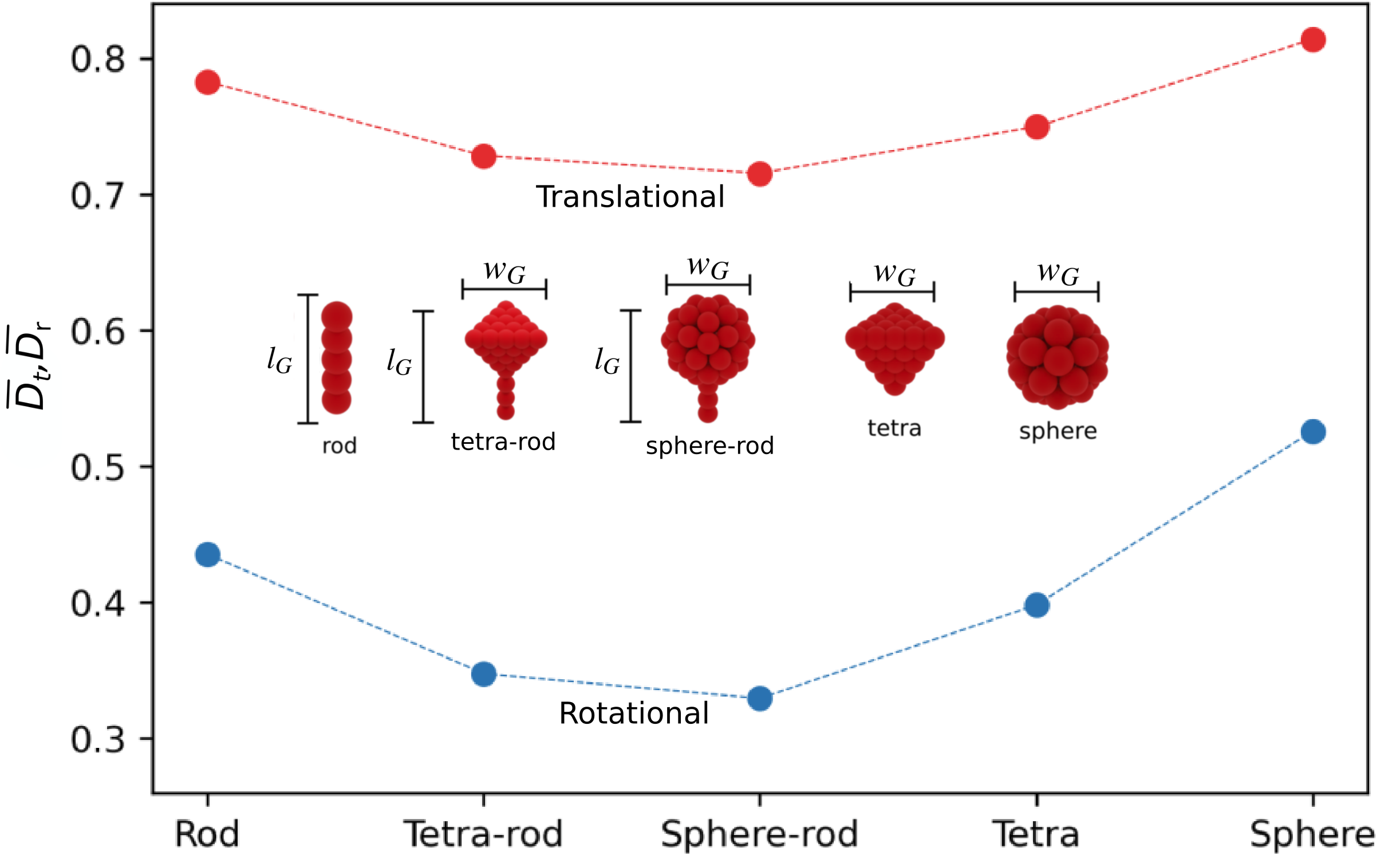}
\caption{Dimensionless rotational and translational diffusivity for different type of groups. }
\end{figure}

\subsubsection{Size of the functional groups} 

As discussed in the previous section, the characteristic size of the groups influences the effective transport of the nanoparticles. Thus, we investigate the changes in \nps\ mobility as the length ($l_G/R$) and/or width ($w_G/R$) of \gs\ varies. In FIG 4, we present  the variation of $\bar{D}_t$ and $\bar{D}_r$ for \nps\ with rod and sphere-rod \g, with $N_G=20$ randomly distributed and different $l_G /R$. For this test, sphere-rod groups have a fixed size of { $w_G/R=0.2$} and the variation in $l_G$ is attained by changing the length of the rod. As presented in FIG 3, compared to simple rods, the sphere-rod groups exhibit the lowest mobility. However, the decay with $l_G/R$ for both types is consistently preserved (see FIG 4) indicating a strong correlation with the overall length of the group. Due to the scaling of the rotational diffusion ($\overline{D}_r \propto R^3$), the increase in the length of the groups induces a significant reduction ranging from 20 to 60 percent. In addition to the group length, we also inspect the effect of group width $w_G /R$. In FIG 5, we compare the diffusional decay for \nps\ with sphere, tetra and sphere-rod groups with three different $w_G /R$. Similar to the previous case, the rate of decay shows to be preserved across the different morphologies. In summary, we identify that regardless of the group morphology the diffusion of the \nps\ scales with the group length in similar fashion. As a consequence, for practical applications although the direct comparison of \np\ diffusivity can differentiate functional groups morphology, the estimation of the diffusional decay can provide a generic indirect measurement of the thickness of the functional groups decorating a nanoparticle. 
 
\begin{figure}[h]
\centering
{\includegraphics[scale = 0.12]{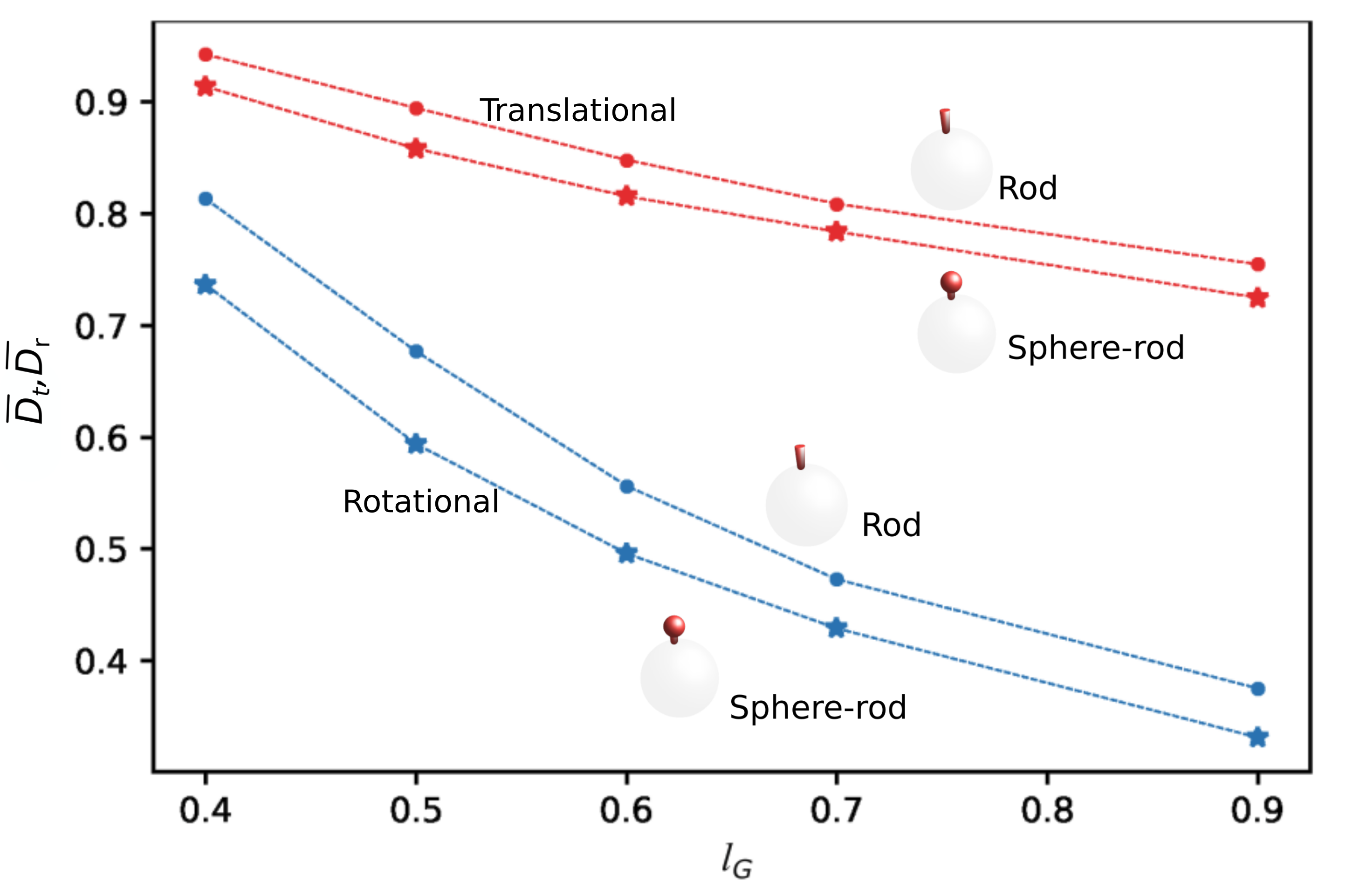}}
\caption{Dimensionless rotational and translational diffusivity for different \g\ length $l_G/R$. Nanoparticles with $N_G$=20  randomly distributed}
\end{figure}
\vspace{-0.5 cm}
\begin{figure}[h]
{\includegraphics[scale = 0.18]{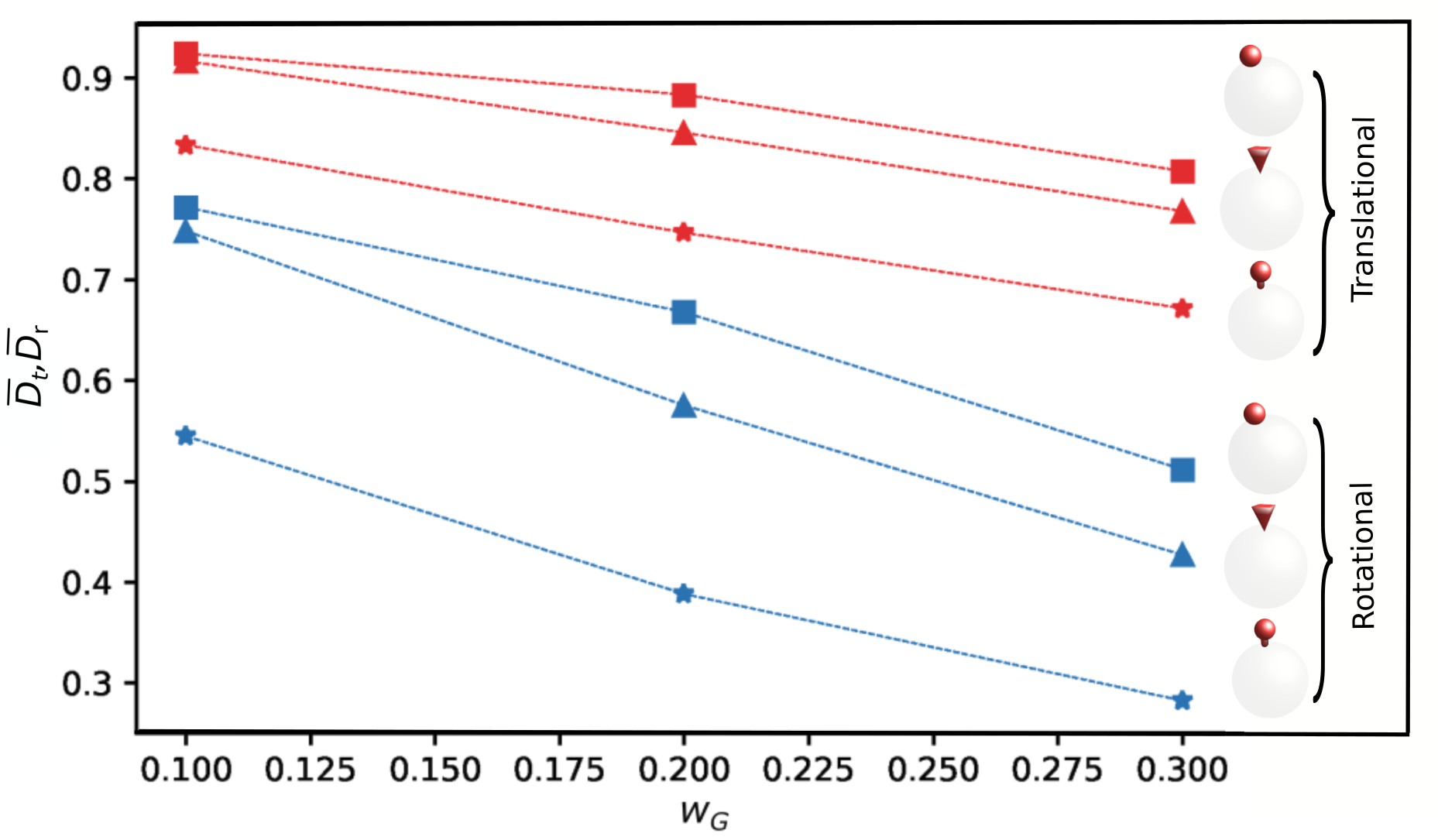}}
\caption{Dimensionless rotational and translational diffusivity for different \g\ width $w_G/R$. Nanoparticles with $N_G$=20  randomly distributed}
\end{figure}
\vspace{-0.5 cm}

\subsection{Number and distribution of functional groups}

\subsubsection{Uniform vs random}

In general, the functional groups can be uniform or randomly distributed surface of the \nps\ core. Therefore, to elucidate the possible effect of \gs\ placement in the nanoparticles, we now investigate this effect using rod-type groups. We compute the mobility for \nps\ with $N_G$ ranging from $12$ to $100$, randomly distributed. For each number of groups ten replicas of randomly distributed groups are simulated and the average diffusion results are compiled in FIG 6. Error bars correspond to the standard deviation. For comparison, in FIG 6, we also include the results for \nps\ with uniform distribution with $N_G$ equal to $12, 42$, and $162$. For simplicity, the uniform distribution of \gs\ is attained by localizing the groups at the vertex of a regular tetrahedron (e.g. $N_G=12$ in a icosahedron). In FIG 6, the results for uniform distribution with $N_G=100$  corresponds to the interpolated value between $N_G=42$ and $N_G=162$.   
\begin{figure}[!ht]\label{randomgraphic}
\centering
{\includegraphics[scale = 0.09]{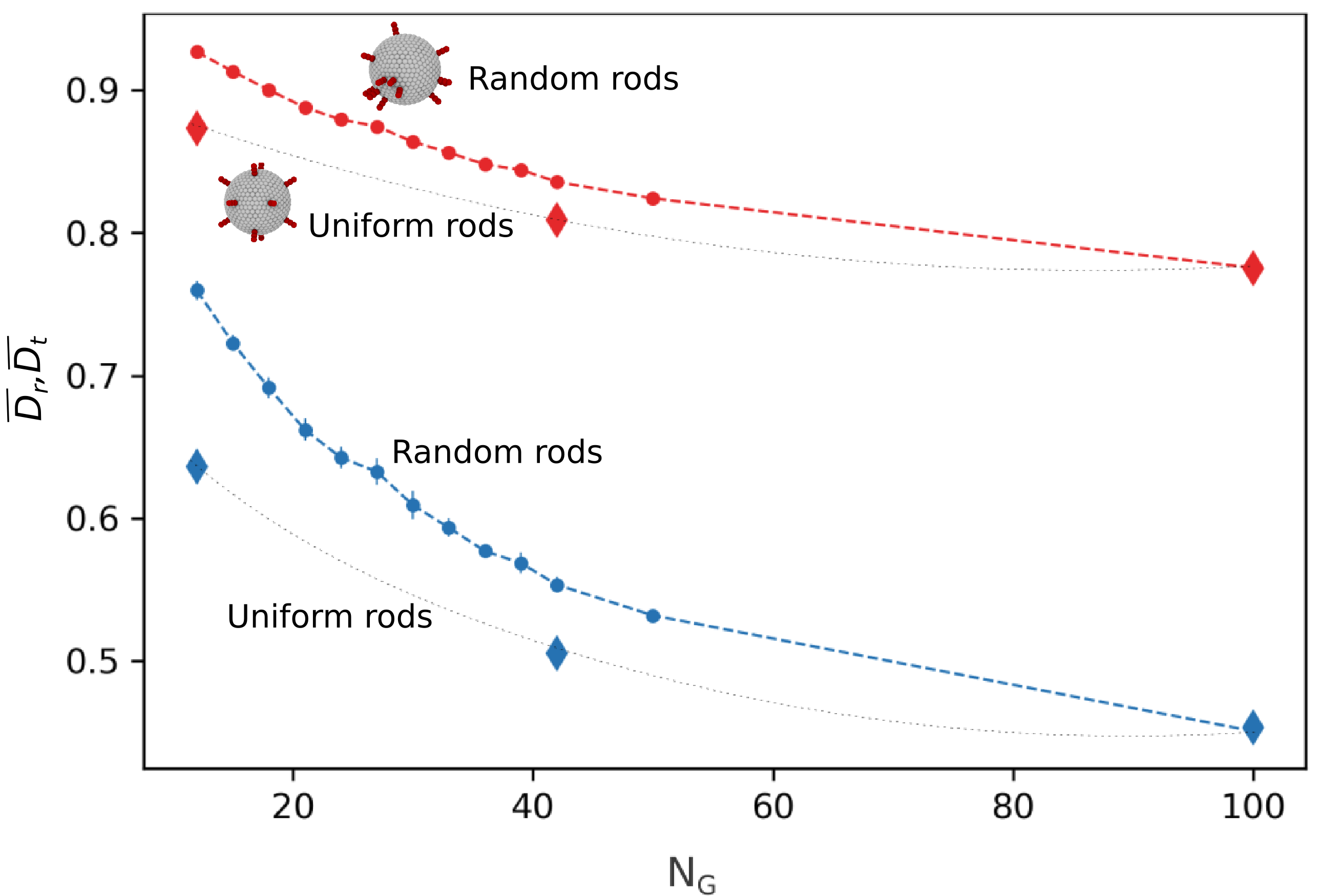}}
\caption{Dimensionless rotational (in blue) and translational (in red) diffusivities with \g\ randomly and uniformly distributed. Error bars depict the standard deviation of the measured mobility. For uniform distribution, we present results for $N_G=[12, 42,162]$, corresponding to the equidistant vertex of regular polyhedrons. The values of 100 \g\ are calculated with interpolation between 42 and 162. However, it is evident that at large $N_G$, randomly and uniformly distribution converges due to the packing of the \g\ in the core.
}
\end{figure}
From FIG 6, we identify that uniform distributions lead to diffusivities of approximately 6$\%$ (for translational) and 16$\%$ (for rotational) smaller than the random ones. The breaking in symmetry of the randomly distributed groups potentially enhance therefore the mobility of the \nps.  If $N_G$ increases, both type of distributions lead to similar effective diffusivities (difference of around 0.07$\%$ for translational and 0.5$\%$ for rotational). At large $N_G$ the hydrodynamic effects of the groups overlap leading to indistinguishable effect of location. We must highlight that the stronger effect on rotational diffusivity at lower groups number is an interesting feature that can be potentially used for nanoparticle characterization and design.

\subsubsection{Number of functional groups}

In the previous section, we already showed that the increase in $N_G$ leads to a reduction of the \nps\ diffusion, reaching a condition where random and uniform are nearly equivalent. Now, we further investigate the effect of groups number on the \np\ mobility. In FIG 7, we compile the diffusivities of \nps\ with a rod, sphere-rod and tetra-rod shape groups for $N_G$ ranging from $10$ to $50$ randomly distributed. The size of the groups is $w_G=0.25, l_G=0.8$ for the tetra-rod, $w_G=0.2, l_G=0.8$ for the sphere-rod, and $l_G=0.8$ for rods. We observe that although the type of \g\ determines the magnitude of the diffusivities, all curves follow similar trend for the three groups. In general, the shape of the group can alter the overall magnitude of the diffusion stemming from the enriched morphological complexity. However, for a fixed group type, it is expected that the group size will be a determinant of the mobility decay with $N_G$. In general, such decay should scale with the characteristic size ($l_G$), but the direct functional relationship cannot be trivially inferred. Therefore, we conduct additional studies considering \nps\ with rod-shape groups of different lengths $l_G/R$ to identify the combined effect of group size and number. In FIG 8, we present the variation in $\overline{D}_t$ and $\overline{D}_r$ of \nps\ with rod-shape groups and three different lengths $l_G/R = 0.5, 0.8, 1.0$. In FIG 8, again a characteristic diffusional decay can be identified, but in this case as $l_G/R$ increases, a stronger dependence with the $N_G$ is elucidated.

\begin{figure}[h]
\centering
\includegraphics[scale = 0.26]{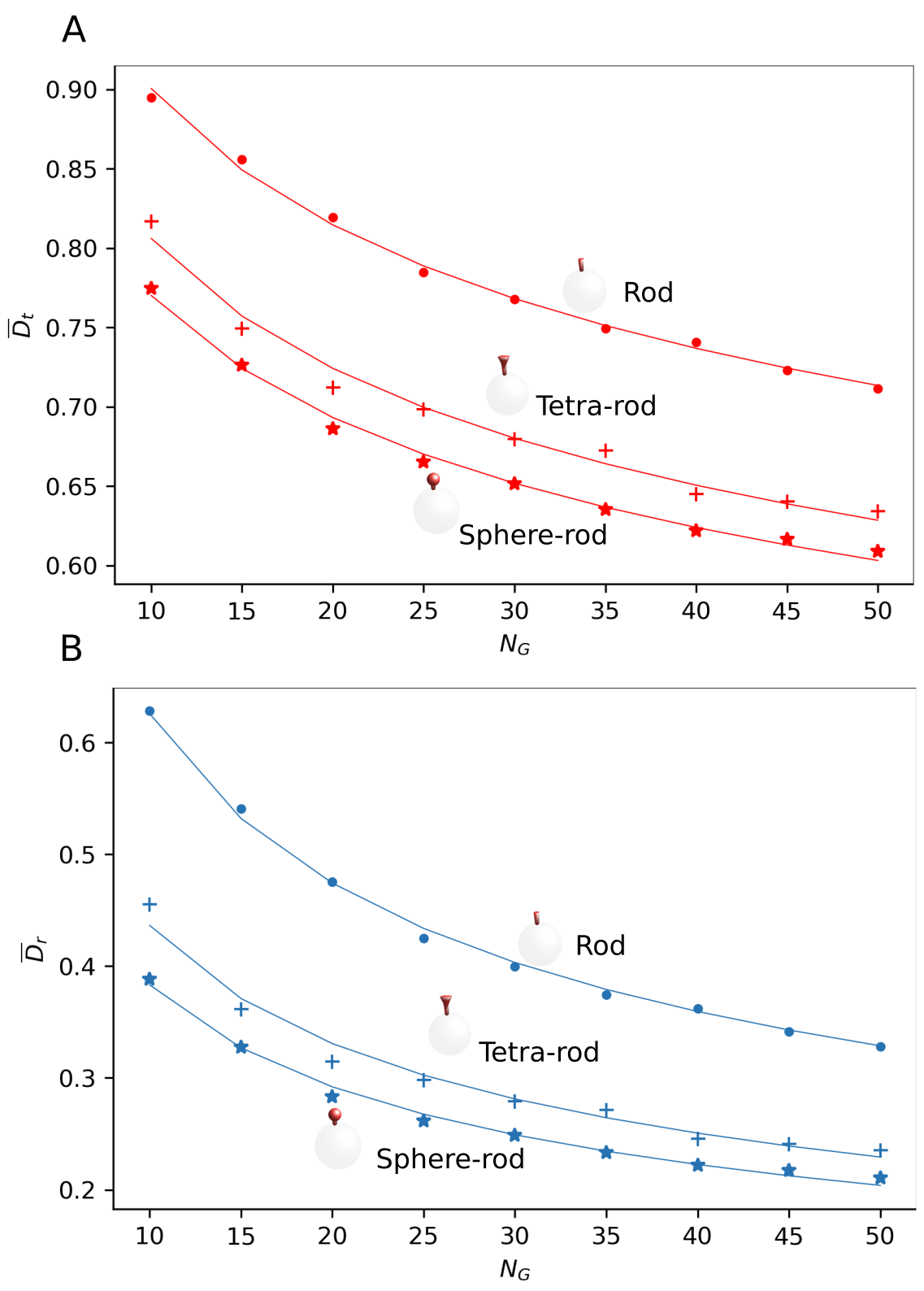}
\caption{Translational and rotational reduced diffusivity of \nps\ with three different types of groups morphology: rods, sphere-rod and tetra-rod. Markers correspond to the computed diffusivities, whereas solid lines indicate the approximated scaling obtained in equations \ref{eq:scalingsT} and \ref{eq:scalingsR}. All groups have $N_G=12$ uniformly distributed, the size of the groups are $w_G=0.25, l_G=0.8$ for the tetra-rod,  $w_G=0.2, l_G=0.8$ sphere-rod, and $l_G=0.8$ for rods}
\end{figure}

\begin{figure}[h]
\centering
\includegraphics[scale = 0.26]{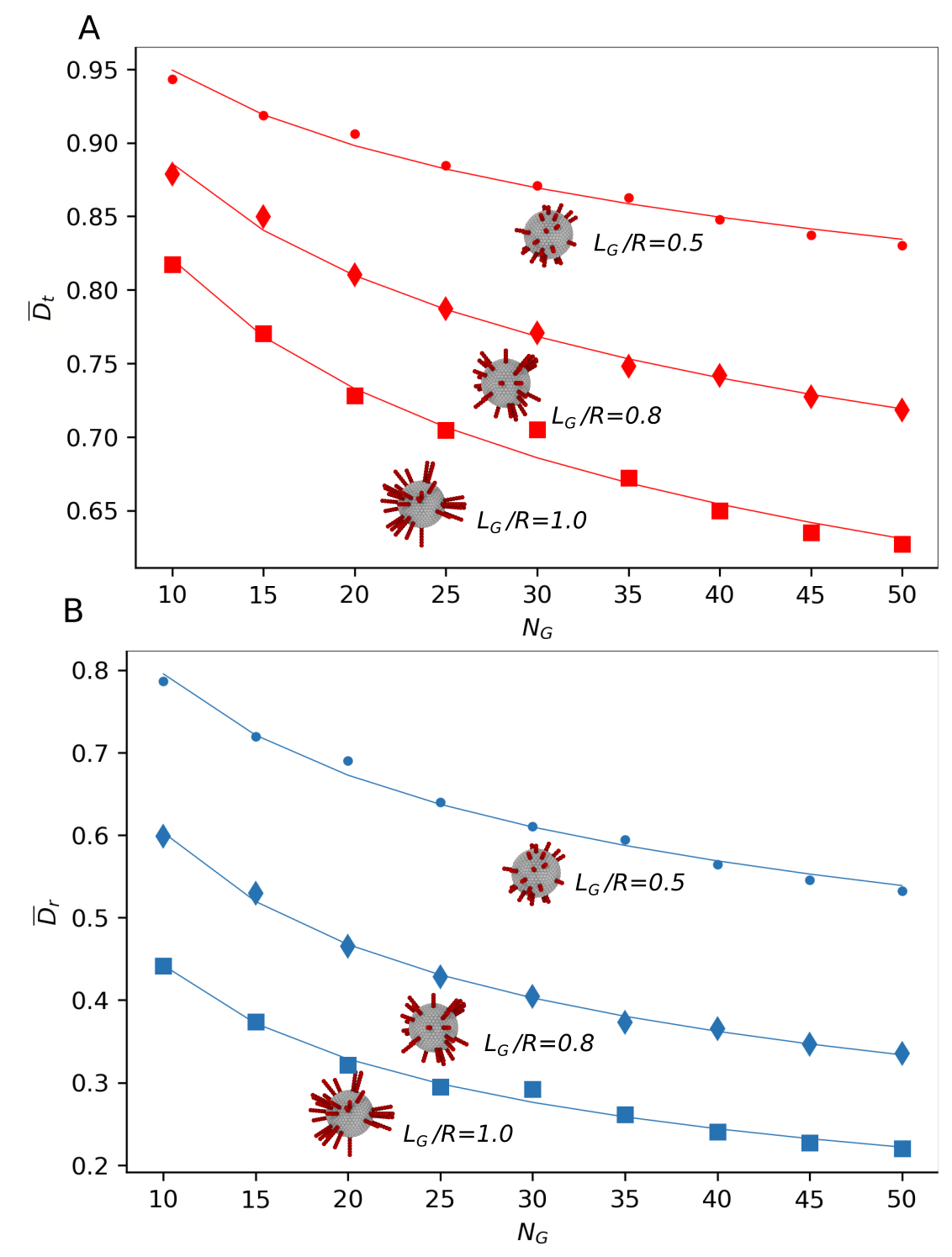}
\caption{A. Translational and B. rotational reduced diffusivity of \nps\ functionalized with rod-shape groups of different lengths $l_G/R = 0.5, 0.8, 1.0$. Markers correspond to the computed diffusivities, whereas solid lines indicate the approximated scaling obtained in equations \ref{eq:scalingsT} and \ref{eq:scalingsR}}
\end{figure}

Stemming from the diffusional decay observed when varying groups type and length (FIG 7 and FIG 8), we can consider a power-law scaling of the diffusion decay as $\overline{D} = 1-\pf N_G^{\sca}$, where $\pf$ and $\sca$ depend on geometrical features ($l_G$ and $w_G$). A numerical analysis on the diffusional decay presented in FIG 7 reveals that the scaling (translational: $\sca \sim -0.08 l_g/R \pm 0.005$ and rotational: $\sca \sim -0.13 l_g/R \pm 0.03$) for the three groups is constant, depending only on the ratio $l_G/R$ that is fixed for all the \g\ types. In contrast, the prefactor $\pf$ that determines the overall magnitude of the decay, depends on the both  $l_G/R$ and $w_G/R$. In  {\color{black} Table III}, we give a break down of the estimated $\pf$ and $\sca$ for each case. The prefactor $\pf$ has a linear Pearson correlation of $r\approx -0.99$ with the volume of the group. Similarly, from the results shown in FIG 8, the estimation of $\pf$ and $\sca$ parameters corroborate their dependency with geometrical features ({\color {black} see Table IV}).  From these results, we can draw the following approximation of the diffusional decay as
\begin{align}
\overline{D}_t - 1 &\propto  v_G N_G^{-0.08l_G},  \label{eq:scalingsT} \\ 
\overline{D}_r - 1 &\propto  v_G N_G^{-0.15l_G},  \label{eq:scalingsR}
\end{align}
where $v_G = {{w_G}^2l_G}/{R^3}$. Overall, the diffusional decay of the functionalized nanoparticles with $N_G$ scales with the length of the functional group, regardless of the particular shape. 
\begin{table}[h]
\centering
\caption{Scaling parameters from three different group-types i.e. rod, tetra-rod and sphere-rod for translational and rotational dimensionless diffusion}
\label{tab:my-table}
\begin{tabular}{@{}lllll@{}}
\toprule
           & \multicolumn{2}{l}{$\overline{D}_t$} & \multicolumn{2}{l}{$\overline{D}_r$} \\ \midrule
Type & $c$                & $v$                 & $c$                & $v$                 \\
Rod        & 2.20             & -0.08             & 2.17             & -0.16             \\
Tetra-rod  & 2.08             & -0.08             & 1.79             & -0.13             \\
Shpere-rod & 2.04             & -0.08             & 1.67             & -0.10             \\ \bottomrule
\end{tabular}
\end{table}

\begin{table}[h]
\centering
\caption{Scaling parameters from three different rod $l_G= 0.5,0.8,1.0$ for translational and rotational dimensionless diffusion}
\label{tab:my-table}
\begin{tabular}{@{}lllll@{}}
\toprule
 & \multicolumn{2}{l}{$\overline{D}_t$} & \multicolumn{2}{l}{$\overline{D}_r$} \\ \midrule

Rod-type &$c$                & $v$                 & $c$                & $v$            \\
$l_G=0.5$   & 2.13             & -0.07             & 2.24             & -0.19             \\
$l_G=0.8$   & 2.15             & -0.07             & 2.07             & -0.14             \\
$l_G=1.0$   & 2.13             & -0.07             & 1.82             & -0.10             \\ \bottomrule
\end{tabular}
\end{table}
\subsection{Diffusion near rigid walls} 

Many applications of nanoparticles involved their transport under confinement or near walls. Under these conditions, \nps\ diffusion can be significantly affected due to the restricted mobility and the asymmetry in the hydrodynamic interactions exerted on the \np. Here, we explore the effect of a stationary wall in the proximity of functionalized \np\ by estimating the parallel and perpendicular diffusion. \nps\ are located at different dimensionless distance $s'=h/R-1$ to the wall (FIG 1.E), where $h$ is the distance from the \np\ centre to the wall, and $R$ is the core radius. In FIG 9, we initially corroborate that for spherical cores without functionalization, $N_G=0$, the RMB discretization provides the correct theoretical translational diffusion (Eqs.12 and 13). The agreement between the computed translational diffusion and the theoretical one evidences the robustness of RMB to model confined \nps.\cite{Sprinkle2017}. A general expression for the theoretical rotational diffusion near walls is not available over a larger range of $s'$.\cite{Swan2007,BalboaUsabiaga2016} Therefore, based on the computed rotational diffusion for plain spheres, we propose an empirical functional form to describe the change in perpendicular and parallel rotational diffusion with the distance of the \np\ to the wall. We propose polynomial fitting function of the form $f(s') = a(s'^{-3}) + b(s'^{-2}) + c(s'^{-1})+1$. Using this approximation (see FIG 9), we obtain that for parallel rotational diffusion, $a=0.0011$, $b=-0.0125$ and $c=-0.0056$, whereas, for perpendicular, the parameters are $a=0.0039$, $b=-0.0421$ and $c=-0.0061$. For comparison, in FIG 9 we compile the single sphere's translational and rotational diffusion with the corresponding theoretical approximation and the semi-empirical numerical fitting. Overall, we observe that for simple spheres the rotational diffusion is less affected by the wall confinement, converging to the unconfined behaviour at shorter distance, $s' \approx 3$.

\begin{figure}[h]
\centering
\includegraphics[scale = 0.18]{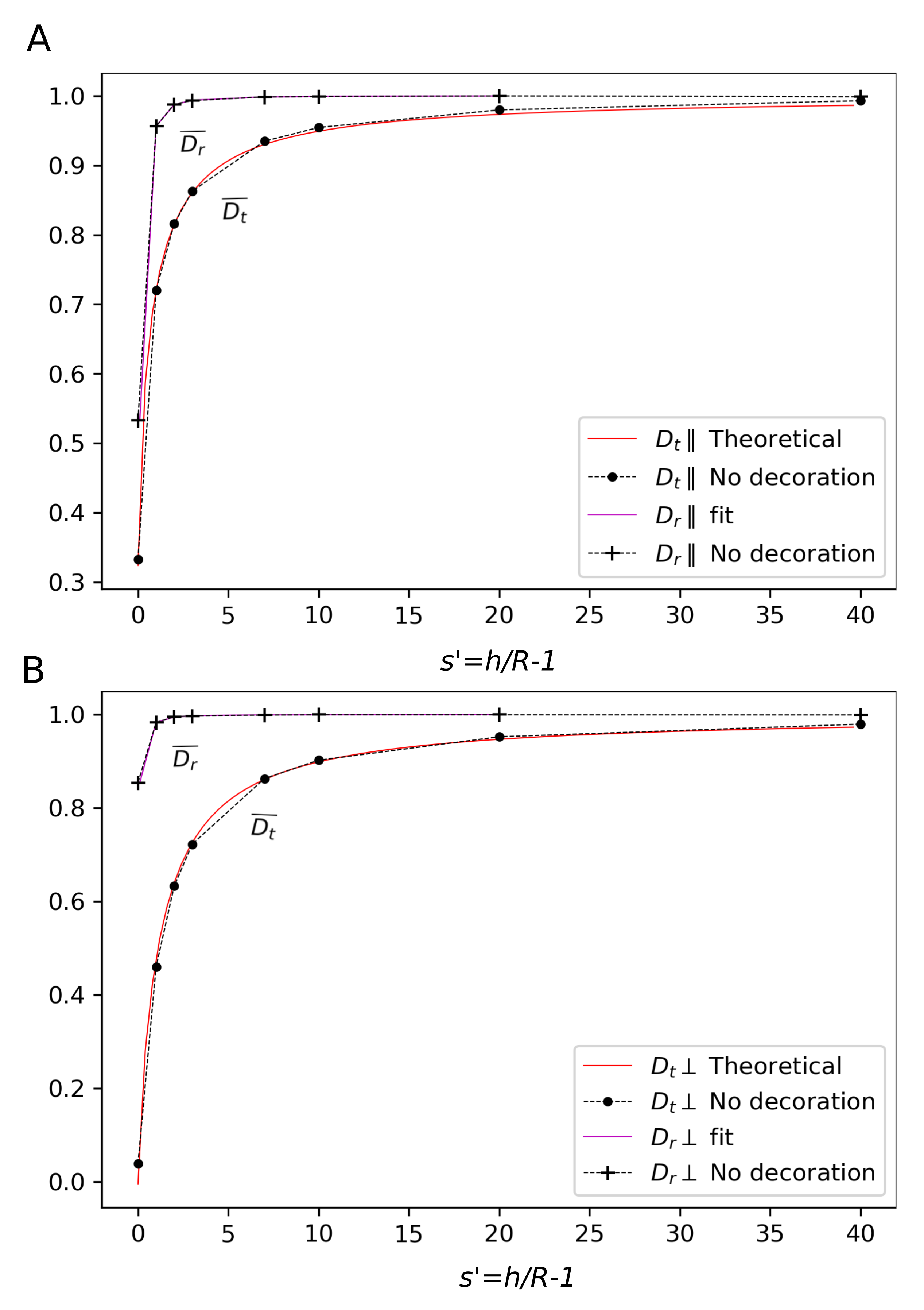}
\caption{A. Perpendicular direction for rotational and translational diffusion. B. Parallel direction for rotational and translational diffusion. Here, we compare rotational and translational according the fitting polynomial function for rotational and the theoretical solution for translational.}
\end{figure}

For confinement studies, we focus on functionalized \nps\ with $N_G=12$ (rod- and tetrahedron-shaped groups) uniformly distributed. 
Simulations for larger $N_G=42$ and randomly distributed groups were also performed for comparison. However, the deviations in those cases were smaller than $1\%$. It is important to note that for the \nps\ near a wall, the RMB method does not have precise lubrication effects. Thus, a reliable distance ($s'$) to capture hydrodynamic effects near the wall requires $s'>(a+l_G)/R$, where $a=0.138$ is the blob radius used in our simulations and $l_G/R=0.5$. We test \nps\ with and without functionalization at $s' = [0.64,1,2,3,7,10,40]$.  In FIG 10, we  present the perpendicular ($D\perp$) and parallel ($D\parallel$) translational diffusivity of the \np. We must recall that the normalization of the diffusion coefficients is done with the corresponding unconfined nanoparticle. Therefore, as the distance to the wall increases, it is expected that the diffusion of the nanoparticles converges to the corresponding unconfined diffusion (i.e  $\overline D_{t}^{wall}\sim 1$). For translational diffusion, this convergence occurs at $s' \sim 40$ (FIG 10), consistent with the theoretical predictions. Our results, indicate that the translational behavior of functionalized \nps\ resembles the non-functionalized ones. Moreover, the change in $\overline D_{t}^{wall}$ for the two types of groups investigated, are practically indistinguishable. Thus, is not expected that changes in translational diffusion can serve as signature of morphological variations, but merely indicate the presence of decorations.  

\begin{figure}[h]
\centering
\includegraphics[scale = 0.17]{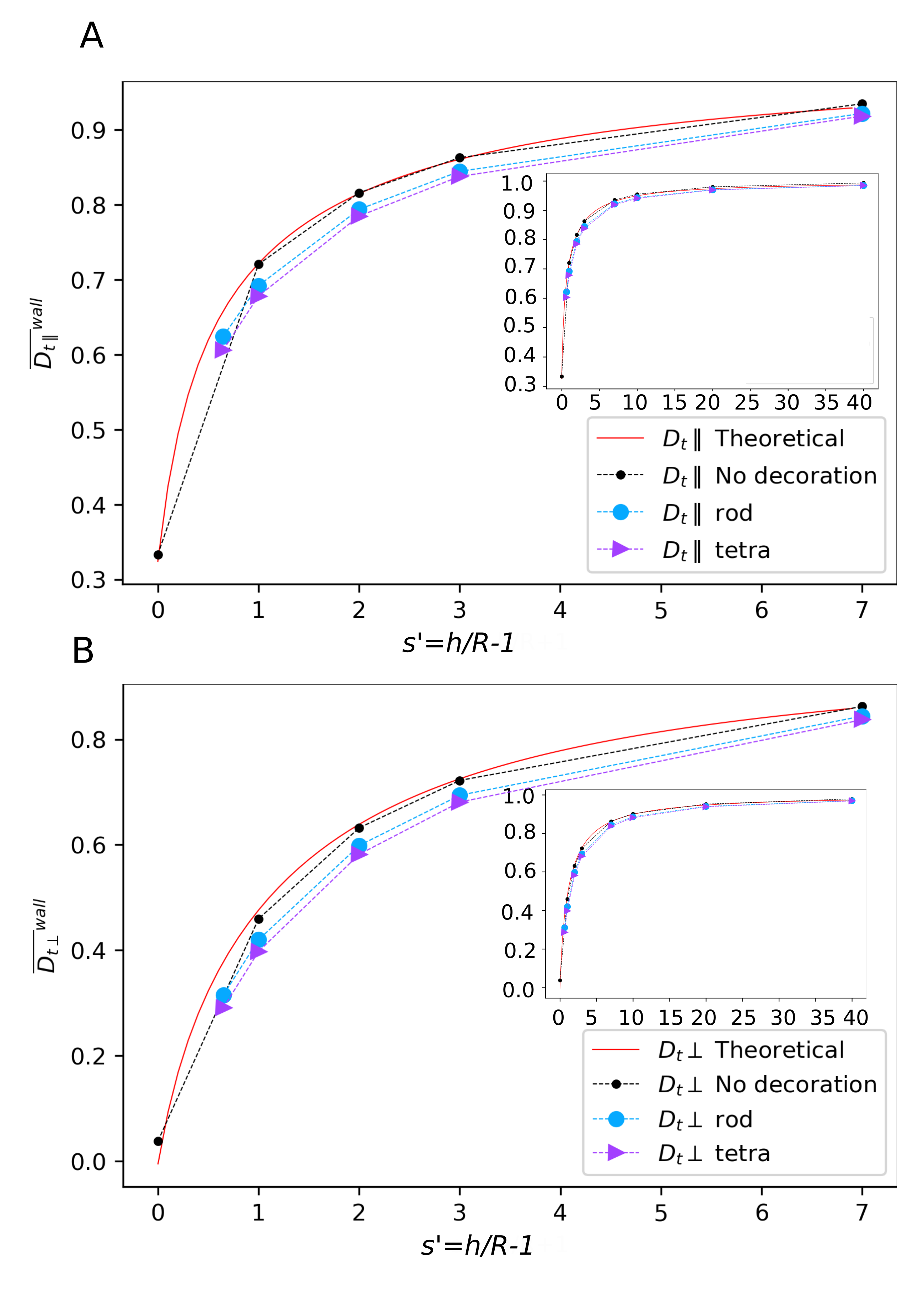}
\caption{A. Parallel  and B. perpendicular translational diffusion for functionalized \np\ with rods and tetra \g, non-functionalized \np, and the theoretical diffusion for spheres.  }
\end{figure}

In FIG 11, we present the confinement effect on the rotational diffusion of the functionalized \nps. Here, we also analyze the diffusion in terms of the parallel and perpendicular component.  In general, the impact of the wall in functionalized and non-functionalized \nps\ conserves a similar trend, where the translational mobility increases more rapidly than the rotational at larger $s'$ distances. However, in the case of functionalized nanoparticles the deviations from the unconfined case are slightly more pronounced. These type of variations indicate that functionalized \nps\ have an enhanced response to confinement, therefore, novel methodologies for characterization, separation, and sensing, can be potentially developed using this characteristic response.

\begin{figure}[h]
\centering
\includegraphics[scale = 0.17]{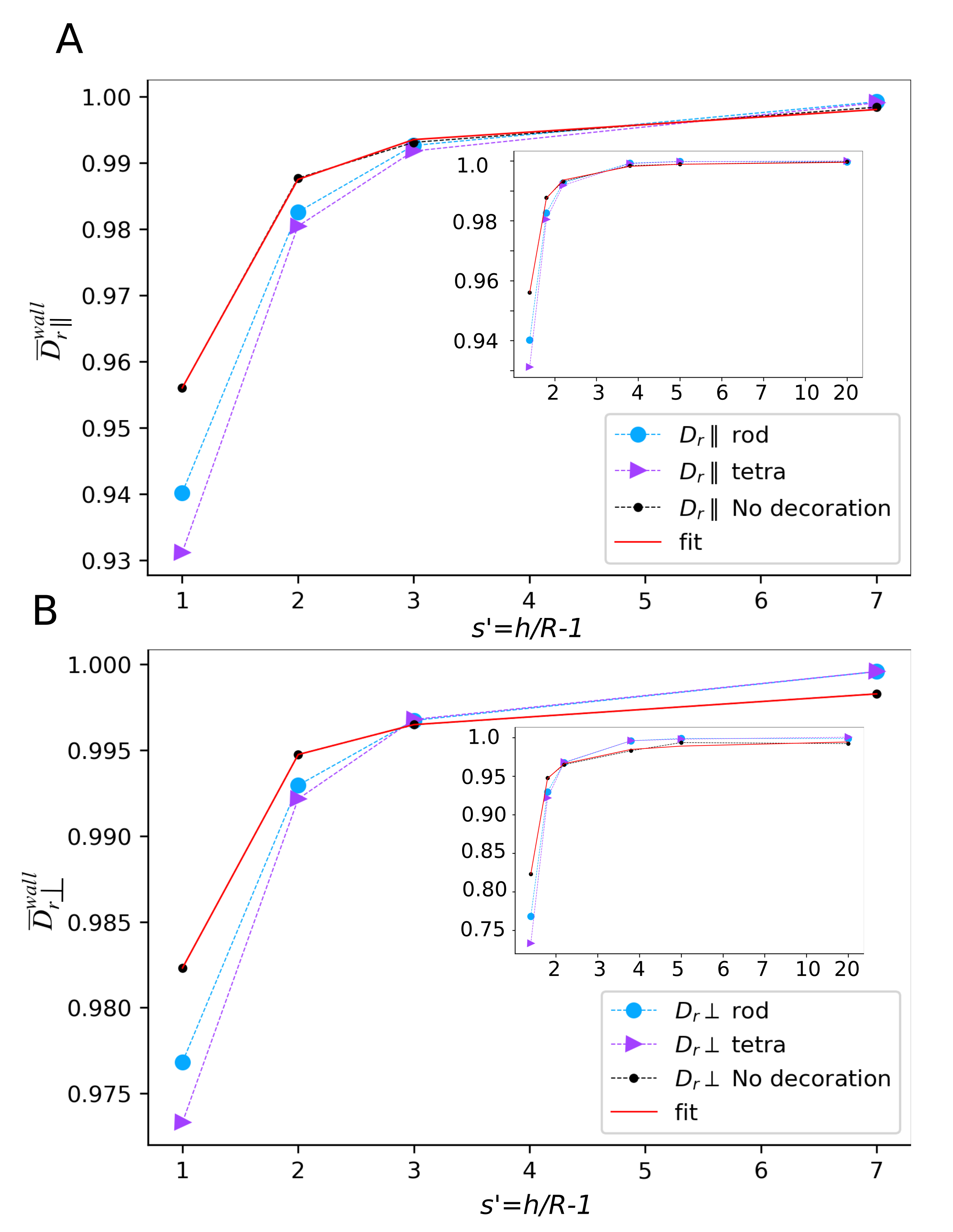}
\caption{Rotational diffusion in A. parallel and B. perpendicular direction from the wall, for different \np\ structures. }
\end{figure}

The source of the slight deviation at small $s'$ evidenced for functionalized nanoparticles in FIG 12 is the combined effect of the break in symmetry due to the wall and the off-diagonal terms in the coupling matrix $\bM_c$ (Eq. 2). For free nanoparticles, these terms tend to zero for the current RMB discretization ($1\text{e}-17$ for non-functionalized and rod-type, and $1\text{e}-09$ for tetra-type). In RMB, the multi-blobs exhibit small off-diagonal components for a sphere but are not zero since the discrete sphere is not perfectly rotational invariant.\cite{BalboaUsabiaga2016} In addition, the resolution ($\Phi$) of the \np\ model interferes with the values of these coupling components; these is attributable to the numerical error. Thus, to analyze the source of the measured off-diagonal terms for functionalized \nps, we use the theoretical mobility of a sphere $M_c^{o}$ with radius $R=1$ as a reference case, $M_c^{o}=1/8\pi \eta R^2$. We study the torque in the $z$ direction, computing the average of the two coupling terms $\bM_{c,xy}$ and $\bM_{c,yx}$, $\bM_{c} = (\bM_{c,xy}-\bM_{c,yx})/2$. Hence, using the off-diagonal terms, we estimate the ratio $M_c/M_c^{o}$, where we compute the difference between the computed coupling matrix from functionalized tetra, rod and the non-functionalized \np. In FIG 12. we observe that the ratio of mobilities increases if the \np\ is near a wall. At a distance, $s' \approx 3$, for the three cases, the coupling terms decay close to zero and similarly to a free \np. The results show that the off-diagonal components for the functionalized \nps\ have higher values than the sphere at short distances to the wall ($s' < 3$). This result evidences that the coupled translational and rotational motion of the functionalized \nps\ near the wall is enhanced. Hence, we identified that, in principle, functionalized nanoparticle characterization could be addressed by placing the \nps\ near walls and measuring the specific variations in the coupling component of the mobility.

\begin{figure}[h]
\centering
\includegraphics[scale = 0.2]{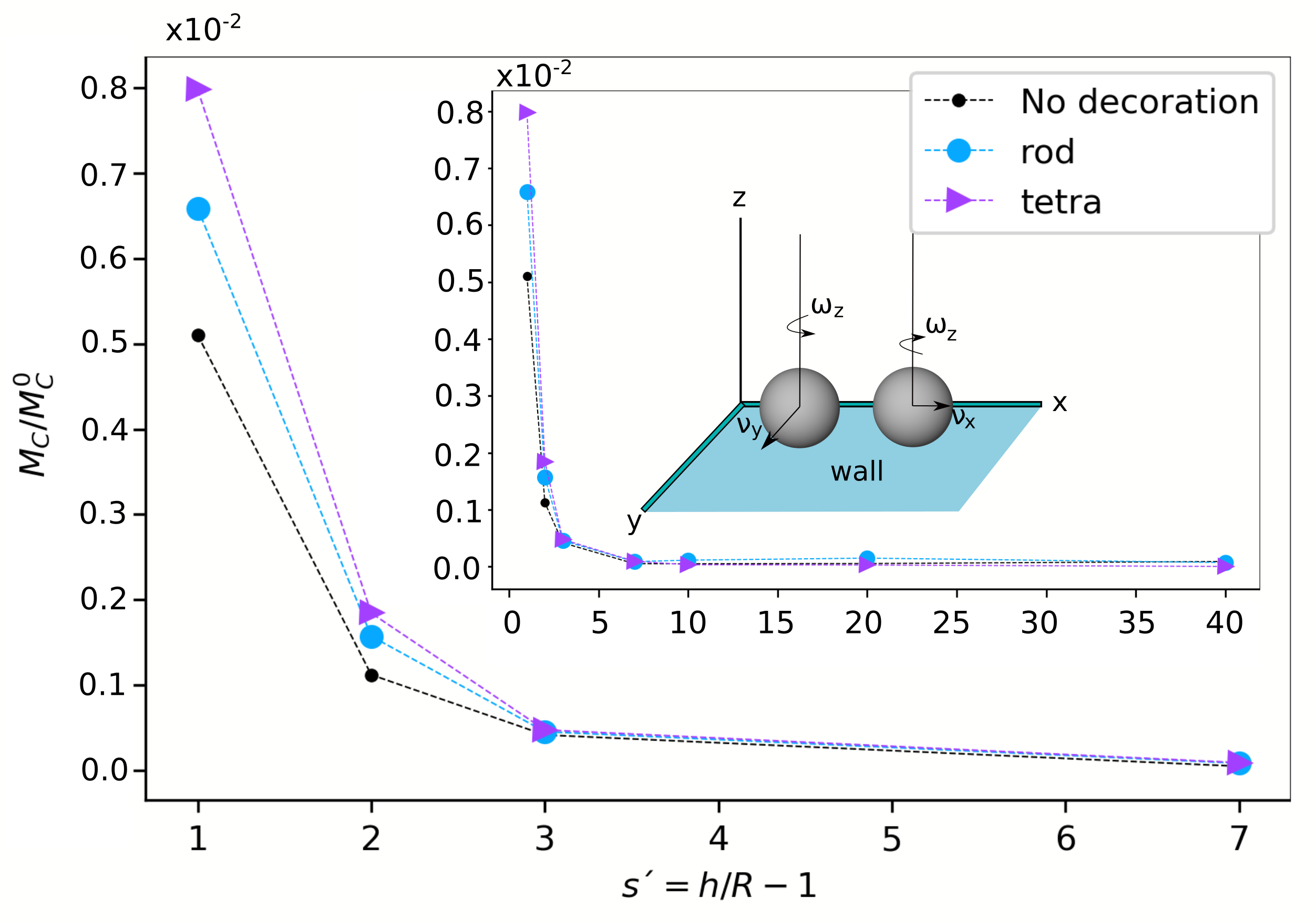}
\caption{Off-diagonal terms of the coupling mobility matrix $\bM_{c}$, these terms are dimensionless with the mobility of a sphere $M_c^{o}$ of radius $R=1$, $ M_c/M_c^{o}$. Starting from the $s' \approx 3$, the mobility converges rapidly to zero, such as a free \np. From values $s'<3$ is evident the translational effects due to torque in $z$ direction} 
\end{figure}

\section{Conclusions}
In this work, we have investigated the passive transport of complex functionalized nanoparticles numerically using a variety of morphologies common in different physical systems. We show how \modi{functional groups} distribution, shape, size, and $N_G$ affect the translational and rotational diffusion of the nanoparticles. In general, we \modi{identify} that the transport properties of the functionalized \nps\ are significantly altered by the morphology of the decorating groups. We observe that functional groups can exhibit a specific reduction in mobility due to added complex morphology of the \nps\. Regarding the group distribution on the \nps\ surface, we identify that random conformations facilitate the transport at a low number of groups compared to uniform distributions due to symmetry-breaking effects. At a large number of groups, the overlap in the hydrodynamic interaction of the groups led to an indistinguishable effect on mobility.

The effect of the number of groups in the diffusion of the nanoparticles exhibits a characteristic power-law decay that is governed by the length of the groups. In contrast, the relative volume determines the overall magnitude of the decay. In general, the characterization of the diffusional decay of functionalized nanoparticles can provide relevant information about the degree of functionalization and size of groups. For \nps\ diffunding near a wall, the diffusion coefficient decay is similar for spherical and functionalized \np. However, functionalized nanoparticles at a short distance from the wall are able to have a more robust response. From a computational standpoint, we show that the RMB method is a powerful tool to characterize and predict the physical transport of complex functionalized \nps. As an additional outcome, we have used RMB to provide a semi-analytical approximation of spherical particles' parallel and perpendicular rotational diffusivity. Our results reveal potential avenues for nanoparticle characterization. We find that confinement effects can be exploited to discern different particle functionalizations, owing to the enhanced response of the rotational diffusion of the \nps\, compared to bulk measurements. Additionally, we showed that targeted modification in the morphology of the groups is a compelling design strategy to create nanoparticles with enhanced or reduced mobility.

\begin{acknowledgments}
Financial support received from the Basque Government through the BERC 2018-2021 program, by the Spanish State Research Agency through BCAM Severo Ochoa excellence accreditation (SEV-2017-0718) and through the project PID2020-117080RB-C55 (“Microscopic foundations of soft-matter experiments: computational nano-hydrodynamics”) funded by AEI - MICIN and acronym “Compu-Nano-Hydro” are  gratefully acknowledged.
N.M acknowledges the support from the European Union’s Horizon 2020 under the Marie Skłodowska-Curie Individual Fellowships grant 101021893, with acronym ViBRheo. F.B.U. acknowledges support from “la Caixa” Foundation (ID 100010434), fellowship LCF/BQ/PI20/11760014, and from the European Union’s Horizon 2020 research and innovation programme under the Marie Sklodowska-Curie grant agreement No 847648.
\end{acknowledgments} 

\section*{Data Availability Statement}
The data that support the findings of this study are available from the corresponding author upon reasonable request.

\nocite{*}
\bibliography{export5}

\end{document}